\documentclass[12pt]{article}

\usepackage{amsfonts}
\usepackage{amssymb}
\usepackage{multirow}
\usepackage{url}

\def\be{\begin{eqnarray}}
\def\ee{\end{eqnarray}}

\begin{document}
\title{Enumeration of plane partitions with a restricted number of parts}
\author{Andrij Rovenchak\\
Department for Theoretical Physics,\\
Ivan Franko National University of Lviv,\\
12, Drahomanov St., Lviv, UA-79005, Ukraine\\
e-mail: andrij.rovenchak@gmail.com}

\maketitle

\begin{abstract}
In the paper, the quantum-statistical approach is used to estimate the number of restricted plane partitions of an integer $n$ with the number of parts not exceeding some finite $N$. The analogy between this number-theoretical problem and the enumeration of microstates of the ideal two-dimensional Bose-gas is used. The conjectured expression for the number of restricted plane partitions shows a good agreement between calculated and exact values for $n=10\div20$.


\end{abstract}

\section{Introduction}
The problem of integer partitions originating yet in works of Leibniz and Euler has found numerous applications not only in mathematics but also in different domains of physics. Mathematical topics include, e.\,g., combinatorics and probability theory, while in physics it is related to theory of crystals, percolation theory or quantum statistics, see, for instance, \cite{Bogoliubov:2007} and references therein.

The so-called two-dimensional or plane partitions are a special type of integer partitions. A plane partition of a positive integer number $n$ is a two-dimensional array of nonnegative integers $n_{ij}$ subject to a nonincreasing condition across rows and columns, such that
\be
n = \sum_{i,j> 0} n_{ij},\qquad\textrm{where}\quad 
n_{i_1j_1}\geq n_{i_2j_2}
\ee
whenever $i_1\leq i_2$, $j_1 \leq j_2$ \cite[p.~176]{Andrews:1976}.
For instance, all the 13 two-dimensional partitions of 4 are \cite{Andrews&Paule:2007}:
{\footnotesize
$$
4,\quad 3\ 1,\quad {3\atop1},\quad 2\ 2,\quad {2\atop2},
\quad 2\ 1\ 1,\quad {2\ 1\atop 1\ \ }, \quad
\begin{array}{c}
2\\ 1\\ 1
\end{array},\quad
1\ 1\ 1\ 1, \quad
\begin{array}{l}
1\ 1\ 1\\ 1
\end{array},\quad
{1\ 1\atop1\ 1},\quad
\begin{array}{l}
1\ 1\\ 1\\ 1
\end{array},\quad
\begin{array}{c}
1\\ 1\\ 1\\ 1
\end{array}.
$$
}

\noindent
Zero elements are traditionally suppressed when writing partitions; remaining non-zero elements are called parts. The number of different plane partitions of $n$ is further denoted as $p^{\rm 2D}(n)$; in the above example $p^{\rm 2D}(4)=13$. In mathematics, this quantity is traditionally referred to as the `partition function'. To avoid ambiguities in this paper, let us retain the term `partition function' for the \textit{Zustandsumme} of the statistical physics and call $p^{\rm 2D}(n)$ the `number of partitions'.

As it is with simple one-dimensional or linear partitions \cite{Auluck&Kothari:1946,Nanda:1951,Grossmann&Holthaus:1997,Tran_etal:2004}, the problem of enumeration of plane partitions can be related to the task of counting the number of microstates in a system of two-dimensional quantum harmonic oscillators obeying the Bose--Einstein statistics \cite{Nanda:1951,Prokhorov&Rovenchak:2012}.

Different types of restrictions might be imposed on partitions. One can consider the parts to be either odd or even numbers, limit the magnitude of parts or their number, etc.\ \cite{Andrews:1976}. With respect to quantum ensembles, this corresponds in particular to studies of fractional statistics or effects of the finite number of particles \cite{Grossmann&Holthaus:1997,Srivatsan_etal:2006,Rovenchak:2009}. With plane partitions, it is also possible to impose different shapes, limit the number of rows and columns, etc. \cite{Andrews:1976,Stanley:1971,Bogoliubov:2005}. Curiously enough, it seems that the problem of enumerating the plane partitions with a sole restriction on the number of parts -- in the sense of asymptotic behavior -- did not find a proper reflection in the literature. The aim of this Letter is to partially fill in this gap.

The paper is organized as follows. Sec.~\ref{sec:General} contains derivation of the relevant expressions for a finite $N$-particle system in the general $D$-dimen\-si\-on\-al case. They are then applied to the 1D problem to recover the known behavior of ordinary (linear) restricted partition in Sec.~\ref{sec:1D}. Restricted plane partitions are considered in Sec.~\ref{sec:2D}. A short discussion concludes the paper.

\section{General results for a finite system of $N$ particles}\label{sec:General}
The partition function $Z_N$ of a finite system of $N$ bosonic harmonic oscillators obey the following recurrence relation \cite{Grossmann&Holthaus:1997,Borrmann&Franke:1993}:
\be\label{eq:ZN-recurr}
Z_N(x) = \frac{1}{N}\sum_{k=1}^N B_{k}(x)Z_{N-k}(x),\qquad 
Z_0(x)\equiv 1,
\ee
where $x={\rm e}^{-\beta\hbar\omega}$ with $\beta$ standing for the inverse temperature and $\omega$ being the oscillator frequency. In $D$ dimensions
\be
B_k(x) = \frac{1}{(1-x^k)^D}.
\ee
A closed-form expression for $Z_N$ exists only in one-dimensional case:
\be\label{eq:ZN1D}
Z^{\rm 1D}_N(x) = \prod_{k=1}^N \frac{1}{1-x^k}.
\ee
We will use this result to check the proposed method.

To solve Eq.~(\ref{eq:ZN-recurr}), and integral transform can be applied. In order to avoid issues of transition from summation to integration, the best choice is seen in using a discrete transform.

The $Z$-transform of a function $f(N)$ is defined as \cite[Chap.~13]{Bracewell:2000}:
\be
\mathfrak{Z}[f(N)] = \sum_{n=0}^\infty f(n) s^{-n} = \tilde f(s).
\ee
It is a discrete analog of the Laplace transform. The following two properties of the $Z$-transform are required to solve Eq.~(\ref{eq:ZN-recurr}): 
\be
\mathfrak{Z}[Nf(N)] = -s \frac{d\tilde f(s)}{ds},
\ee

\be
\mathfrak{Z}[f(N)*g(N)] = \tilde f(s) \tilde g(s),
\ee
where the convolution is defined as:
\be
f(N)*g(N) = \sum_{n=0}^N f(n) g(N-n).
\ee

Let us rewrite Eq.~(\ref{eq:ZN-recurr}) in the form immediately suitable for the application of the $Z$-transform:
\be\label{eq:ZN-recurr1}
N Z_N(x) = \sum_{k=0}^N B_k(x)Z_{N-k}(x),\qquad B_0(x)\stackrel{\rm def}{=}0.
\ee

It seems more convenient to consider the correction to the partition function of an infinite system $Z_\infty(x)$:
\be\label{eq:ZN=Zy}
Z_N(x) = Z_\infty(x) y_N(x),
\ee
where the function $y_N(x)$ has an obvious limiting behavior:
\be\label{eq:yN-lim}
\lim_{N\to\infty}y_N(x) = 1.
\ee
For the transform of this correction we easily obtain:
\be\label{eq:y(s)}
-s \frac{d\tilde y(s|x)}{ds} = \tilde B(s|x) \tilde y(s|x),
\ee
or
\be
\tilde y(s|x) = C\exp\left\{-\int^s \frac{\tilde B(s')}{s'}\,ds'\right\},
\ee
where the integration constant $C$ can be found from Eq.~(\ref{eq:yN-lim}).

\section{Testing the approach in one dimension}\label{sec:1D}
First we check the method for the 1D case where all the results are well known \cite{Grossmann&Holthaus:1997,Tran_etal:2004,Weiss&Holthaus:2002}. 

The summation in the transform of $B_N(x)$ is easily done in the first order of $x$:
\be
\tilde B^{\rm 1D}(s|x) = \sum_{k=1}^\infty \frac{s^{-1}}{1-x^k}\simeq
 \sum_{k=1}^\infty s^{-1}(1+x^k) = \frac{s-2x+sx}{(s-1)(s-x)}
\ee
giving
\be
\tilde y^{\rm 1D}(s|x) = C\frac{s^2}{(s-1)(s-x)}
\ee
and, therefore, by inverting the transform we obtain:
\be\label{eq:yN-Zinverse}
y^{\rm 1D}_N(x) = \mathfrak{Z}^{-1}\left[C\frac{s^2}{(s-1)(s-x)}\right]=
C\frac{x^{N+1}-1}{x-1}.
\ee
From Eq.~(\ref{eq:yN-lim}) the integration constant is $C = 1-x$ and finally in the leading order we have:
\be
y^{\rm 1D}_N(x) = 1 - x^{N+1}.
\ee

This result correctly reproduces exact expression (\ref{eq:ZN1D}). Indeed,
\be
Z^{\rm 1D}_\infty(x) = \prod_{k=1}^N \frac{1}{1-x^k}\prod_{k=N+1}^\infty \frac{1}{1-x^k},
\ee
that is
\be
y^{\rm 1D}_N(x) = \prod_{k=N+1}^\infty (1-x^k) = 
\exp\sum_{k=N+1}^\infty \ln\left(1-x^k\right).
\ee
Again in the leading order, taking into account that $x<1$ and $N$ is large, we obtain:
\be\label{eq:yN-exact}
y^{\rm 1D}_N(x) = \exp\left(-x^{N+1}\right) = 1-x^{N+1}\pm\ldots.
\ee

At this point some clarifications are required. The number of (one-dimensional or linear) partitions $p^{\rm 1D}(n)\equiv p(n)$ of an integer $n$ equals to the number of microstates ${\it\Gamma}(E)$ of the systems with energy $E=\hbar\omega n$. Function ${\it\Gamma}(E)$ is linked to $Z(\beta)$ via the inverse Laplace transform, which can be evaluated using the method of the steepest descent as follows \cite{Tran_etal:2004}:
\be
{\it\Gamma}(E) = \frac{{\rm e}^{S(\beta_0)}}{\sqrt{2\pi S''(\beta_0)}},
\ee
where the entropy $S(\beta) = \beta E + \ln Z(\beta)$, and the stationary point $\beta_0$ is defined by $S'(\beta_0)=0$.

Considering a finite number of particles $N$ (or equivalently, a finite number of parts for the partitions) one can see from Eq.~(\ref{eq:ZN=Zy}) that expression for $y_N(x)$ should enter directly into the formula for the number of restricted partitions:
\be
p_N(n) = p(n) y_N({\rm e}^{-\beta_0}),
\ee
where the stationary point is $\beta_0=\pi/\sqrt{6n}$ \cite{Tran_etal:2004,Prokhorov&Rovenchak:2012}. Note that as $n$ is large, $\beta_0$ is a small number and the argument $x={\rm e}^{-\beta_0}$ is close to unity. This results in a small modification in (\ref{eq:yN-exact}), namely:
\be\label{eq:yN-exact1}
y_N(x) = \exp\left(-\frac{x^{N}}{1-x}\right) \simeq  \exp\left(-\frac{x^{N}}{\beta_0}\right).
\ee
Therefore, the leading correction in the number of restricted partitions is \cite{Auluck&Kothari:1946,Tran_etal:2004}
\be
p_N(n) = p(n) \exp\left\{-\frac{\sqrt{6n}}{\pi}\,{\rm e}^{-\pi N/\sqrt{6n}}\right\},
\ee
reproducing the classical result of Erd\H{o}s and Lehner \cite{Erdos&Lehner:1941} about the asymptotic behavior of the number of partitions of $n$ into at most $N$ parts.

Obtaining such an expression directly from Eq.~(\ref{eq:yN-Zinverse}) would be somewhat speculative as the $Z$-transformed function $\tilde B^{\rm 1D}(s)$ was derived in the limit of small $x$. It is possible to solve in a closed form the equation for $\tilde y^{\rm 1D}(s)$ in the limit of $x\to 1$ but the inversion $\mathfrak{Z}^{-1}[\tilde y^{\rm 1D}(s)]$ cannot be made analytically in this case. So, one can compare -- with respect to $N$ and $x$ -- the \textit{structure} of expressions for $y^{\rm 1D}_N(x)$ from the two approaches and assume such comparison to be valid in higher space dimensions as well. As is shown below, this assumption leads to a quite good agreement between real and calculated numbers of restricted plane partitions.

\section{Results for plane partitions}\label{sec:2D}
In two dimensions, $B^{\rm 2D}_k(x)$ equals:
\be
B^{\rm 2D}_k(x) = \frac{1}{(1-x^k)^2},
\ee
giving approximately
\be
\tilde B^{\rm 2D}(s|x) \simeq  \sum_{k=1}^\infty s^{-1}(1+2 x^k) = \frac{s-3x+2sx}{(s-1)(s-x)}.
\ee
The solution of Eq.~(\ref{eq:y(s)}) is
\be
\tilde y^{\rm 2D}(s|x) = C\frac{s^3}{(s-1)(s-x)^2}
\ee
and its inverse $Z$-transform reads:
\be
y^{\rm 2D}_N(x) = \mathfrak{Z}^{-1} [\tilde y^{\rm 2D}(s|x)] = 
C\frac{(N+1)x^{N+2}-(N+2)x^{N+1}+1}{(x-1)^2}
\ee
with $C=(x-1)^2$, becoming for large $N$ and small $x$:
\be\label{eq:yN-2D}
y^{\rm 2D}_N(x) = {1-N x^{N}}
\ee

A similar expression can be obtained for a system of $N$ isotropic two-dimensional oscillators if the partition function is written in the form:
\be
\ln Z^{\rm 2D}_N(x) = -\sum_{k=1}^N k\ln(1-x^k),
\ee
where the $k$-fold degeneracy of the $k$th level is taken into account. With the upper limit extended to infinity this recovers MacMahon's generation function for plane partitions \cite{MacMahon:1897}:
\be
\sum_{n=0}^\infty p^{\rm 2D}(n) x^n = \prod_{n=0}^\infty \frac{1}{(1-x^n)^n}.
\ee

From the structure of (\ref{eq:yN-2D}), acting by analogy with the one-dimensional case to consider the $\beta\to0$ limit, we can assume the following asymptotics for $y^{\rm 2D}_N(x)$:
\be
y^{\rm 2D}_N(x) = \exp\left(-\frac{N x^N}{(x-1)^2}\right) = \exp\left(-\frac{N x^N}{\beta_0^2}\right),
\ee
where the stationary point 
\be
\beta_0 = \left(\frac{2\zeta(3)}{n}\right)^{1/3}
\ee
with $\zeta(x)$ standing for Riemann's zeta-function \cite{Prokhorov&Rovenchak:2012}.

Therefore, the following asymptotic behavior can be conjectured for the number of restricted plane partitions:
\be\label{eq:MAIN}
p^{\rm 2D}_N(n) = p^{\rm 2D}(n) 
\exp\left\{-\frac{N n^{1/3}}{[2\zeta(3)]^{1/3}} {\rm e}^{-N [2\zeta(3) / n]^{1/3}}\right\}.
\ee
This formula is the main result of the paper. It gives the estimation of the number of plane partitions of $n$ into at most $N$ parts.

Conditions for the number of parts $N$ immediately follow from the derivation process:
\be
0.75 n^{1/3} \lesssim N < n,
\ee
where 0.75 is nothing but the approximate value of $[2\zeta(3)]^{-1/3}$.

Some calculation results with Eq.~(\ref{eq:MAIN}) are given in Table~\ref{tab:results}. Exact numbers of unrestricted plane partitions $p^{\rm 2D}(n)$ can be found in \cite{oeis:PLn}, while the asymptotic dependence on $n$ is given by \cite{Wright:1931,Mutafchiev&Kamenov:2006}
\be\label{eq:p2D}
p^{\rm 2D}(n) = \frac{[2\zeta(3)]^{7/36}}{\sqrt{6\pi}}\, n^{-25/36}
\exp\left\{\frac32 [2\zeta(3)]^{1/3} n^{2/3} + c\right\},
\ee
where $c = \zeta'(-1) = -0.165421\ldots\ $. In \cite{Prokhorov&Rovenchak:2012} the value of $c=-1/6=-0.166666\ldots$ was obtained which gives a better approximation for $n \leq 7573$ but fails to catch the correct behavior for $n\to\infty$.

\bigskip
\begin{table}[h]
\begin{center}
\begin{tabular}{cccccccccc}
\hline

\multirow{2}{*}{$n,N$} & 
\multirow{2}{*}{$p^{\rm 2D}(n)$} & 
\multicolumn{4}{c}{$p^{\rm 2D}_N(n)\vphantom{\int^N}$} & 
\multicolumn{3}{c}{relative} \\
\cline{3-6}
         &                         & exact & calc.\,1 & calc.\,2 & calc.\,3 & \multicolumn{3}{c}{errors, \%} \\
\hline
$n=10,\ \ N=\phantom{0}9$   & 500  & 458  & 474 & 498 & 497 &
                                               3.5 & 8.8 & 8.7 \\
$n=15,\ \ N=14$ & 6879 & 6703 & 6791 & 7082 & 7073 & 
                                       1.3 & 5.7 & 5.5 \\
$n=20,\ \ N=19$ & 75278 & 74651 & 75003 & 77574 & 77478 &
                                       0.5 & 3.9 & 3.8 \\
$\phantom{n=20,\ \ }
              N=18$ &           & 74161 & 74898 & 77435 & 77339 &
                                        1.0 & 4.4 & 4.3\\
\hline
\end{tabular}
\end{center}
\caption{Number of restricted plane partitions. Exact values are taken from \cite{oeis:PLnN}. The `calc.\,1' column corresponds to exact $p^{\rm 2D}(n)$ from \cite{oeis:PLn}; 
the `calc.\,2' and `calc.\,3' columns are based on $p^{\rm 2D}(n)$ from (\ref{eq:p2D}) with the constant $c$ taken from \cite{Prokhorov&Rovenchak:2012} and \cite{Wright:1931,Mutafchiev&Kamenov:2006}, respectively.}\label{tab:results}
\end{table}

\section{Discussion}\label{sec:Discussion}
As seen from Table~\ref{tab:results}, a good accuracy is achieved for the number of restricted plane partitions given by Eq.~(\ref{eq:MAIN}), at least for $n=10\div20$. Higher relative errors (4--9 per cent) are mainly due to the error of the asymptotic formula (\ref{eq:p2D}) for the number of unrestricted plane partitions. An expected monotonous decrease of the relative error is observed at least for $N/n = 9/10$.

Further planned studies of this problem include numerical solution of the recurrence equation (\ref{eq:ZN-recurr}) for $Z_N(x)$ to check the $x$ and $N$ behavior, especially in the limits of small $x$ and large $N$. Once exact values of $p^{\rm2D}_N(n)$ are available, the conjectured formula (\ref{eq:MAIN}) can be verified for larger $n$.

It seems tempting to extend the suggested approach onto higher-dimen\-si\-on\-al partitions. However, I will refrain from doing so before a thorough analysis of restricted plane partitions is made.

\section*{Acknowledgement}
The work was partly supported by grant $\Phi\Phi$-110$\Phi$ (No. 0112U001275) from the Ministry of education and science of Ukraine.


\begin{thebibliography}{00}
\bibitem{Bogoliubov:2007}Bogoliubov N.~M.,
Theor. Math. Phys. \textbf{150} (2007) 165.

\bibitem{Andrews:1976}Andrews G.~E., \textit{The Theory of Partitions} (Addison-Wesley, Reading, Mass.) 1976.

\bibitem{Andrews&Paule:2007}Andrews G. and Paule P., 
J. London Math. Soc. (2) \textbf{76} (2007) 647.

\bibitem{Auluck&Kothari:1946}Auluck F. C. and Kothari D. S., 
Proc. Camb. Phil. Soc. \textbf{42} (1946) 272.

\bibitem{Nanda:1951}Nanda V. S.,  
Proc. Camb. Phil. Soc. \textbf{47} (1951) 591.

\bibitem{Grossmann&Holthaus:1997}Grossmann S. and Holthaus M., 
Phys. Rev. Lett. \textbf{79} (1997) 3557.

\bibitem{Tran_etal:2004}Tran M. N., Murthy M. V. N. and Bhaduri R. J., 
Ann. Phys. \textbf{311} (2004) 204.

\bibitem{Prokhorov&Rovenchak:2012}Prokhorov D. and Rovenchak A.,
Condens. Matter Phys. \textbf{15} (2012) 33001.

\bibitem{Srivatsan_etal:2006}Srivatsan C. S., Murthy M. V. N., and  Bhaduri R.K., Pramana -- J. Phys. \textbf{66} (2006) 485.

\bibitem{Rovenchak:2009}Rovenchak A., 
Fiz. Nizk. Temp. \textbf{35} (2009) 510; Low Temp. Phys. \textbf{35} (2009) 400.

\bibitem{Stanley:1971}Stanley R. P.,
Stud. Appl. Math. \textbf{50} (1971) 167, 259.

\bibitem{Bogoliubov:2005}Bogoliubov N. M.,
J. Phys. A: Math. Gen. \textbf{38} (2005) 9415.

\bibitem{Borrmann&Franke:1993}Borrmann P. and Franke G.,
J. Chem. Phys. \textbf{98} (1993) 2484.

\bibitem{Bracewell:2000}Bracewell R. N.,
\textit{The Fourier transform and its application}, 3rd edition (McGrow Hill, 2000).

\bibitem{Weiss&Holthaus:2002}Weiss C. and Holthaus M.,
Europhys. Lett. \textbf{59} (2002) 486.

\bibitem{MacMahon:1897}MacMahon P. A., 
Phil. Trans. R. Soc. London A \textbf{187} (1897) 619.

\bibitem{Erdos&Lehner:1941}Erd\"os P. and Lehner J.,
Duke Math. J. \textbf{8} (1941) 335.

\bibitem{Wright:1931}Wright E. M., 
Q. J. Math. \textbf{os-2} (1931) 177.

\bibitem{Mutafchiev&Kamenov:2006}Mutafchiev L. and Kamenov E., Compt. Rend. Acad. Bulg. Sci. \textbf{59} (2006) 361.

\bibitem{oeis:PLn}Number of planar partitions of n, \textit{On-line Encyclopedia of Integer Sequences}, \url{http://oeis.org/A000219}.

\bibitem{oeis:PLnN}Triangle read by rows: T[n,k]= plane partitions of n containing k parts, \textit{On-line Encyclopedia of Integer Sequences}, \url{http://oeis.org/A091298}.

\end{thebibliography}
\end{document}